\documentclass[journal]{IEEEtran}

\usepackage{graphicx} 
\usepackage{booktabs}
\usepackage{color}
\usepackage{comment}

\begin{document}

\title{Label-Preserving Fourier Domain Adaptation with Generative-AI and Real Style Donors for 3D Mitral Valve Segmentation}
\author{Javeria Javed, Ahmed Salim, Muhammad Mahboob Ur Rahman~\IEEEmembership{Senior Member,~IEEE}
\thanks{Javeria Javed is with the Department of Computer and Software Engineering, College of Electrical and Mechanical Engineering (CEME), NUST, Rawalpindi, Pakistan. Ahmed Salim is with the Electrical Engineering Department, Namal University, Mianwali, Pakistan. Muhammad Mahboob Ur Rahman is with the Electrical Engineering Department, Information Technology University (ITU), Lahore, Pakistan.}

}

\maketitle

\begin{abstract}

% v3 (latest)

Accurate segmentation of the mitral valve from three-dimensional transesophageal echocardiography (3D TEE) is severely restricted by the scarcity of publicly available expert-annotated volumetric datasets. To address this critical limitation, we investigated Fourier Domain Adaptation (FDA), utilizing general-purpose generative-AI images (from platforms like ChatGPT, Google Gemini, and Microsoft MAI) as annotation-free style donors for 3D mitral valve segmentation.
We employed a DeepMitral residual U-Net model on the public MVSeg2023 dataset, comparing six conditions over three runs: real-only training, traditional augmentation, FDA with each generative source, and a non-AI FDA control using real-image style donors, plus an optimization-matched real-only control and a $\beta$ sensitivity analysis. Case-level paired testing showed that FDA with these generative sources led to significant reductions in boundary- and surface-related errors across all three donor conditions relative to the primary real-only baseline.
However, the impact on the core segmentation metric, Generalized Dice Score (GDS), proved generator-dependent. ChatGPT-FDA showed a significant GDS advantage over the optimization-matched control but not over the primary baseline, and Gemini-FDA showed no significant GDS difference. Notably, only MAI-FDA provided a statistically robust and significant increase in GDS compared to the primary baseline. The $\beta$ sensitivity was not robust across training runs.
These findings confirm that FDA remains an important augmentation mechanism. While generative-AI images offer a practical means of acquiring style diversity without new annotations, the non-AI control performed comparably to ChatGPT-FDA, so we do not conclude that this approach is superior to training with real-image donors.

\end{abstract}


\section{Introduction}

Mitral valve disease, including degenerative mitral regurgitation, functional regurgitation, and rheumatic stenosis, is among the most common structural cardiac conditions requiring surgical or transcatheter intervention~\cite{benjamin2018heart,patwa2025heart}. Three-dimensional transesophageal echocardiography (3D TEE) is the clinical reference modality for assessing mitral valve morphology because it captures the dynamic, saddle-shaped leaflet geometry that two-dimensional imaging cannot fully resolve~\cite{hahn2013guidelines}. Extracting a case-specific mitral valve surface model from a 3D TEE volume is essential for pre-operative planning, in-silico repair simulation, and quantitative assessment of leaflet coaptation, but manual segmentation of a single 3D TEE volume can take a trained operator upwards of two hours~\cite{carnahan2021deepmitral}, making it impractical for routine clinical workflows.

Deep learning has substantially reduced this burden. Carnahan et al.\ introduced DeepMitral, a fully automatic 3D residual U-Net that segments the mitral leaflets directly from raw TEE volumes in seconds rather than hours~\cite{carnahan2021deepmitral}, and the same group subsequently released the associated data through the MVSeg2023 public challenge~\cite{mvseg2023}. Despite this progress, the fundamental data bottleneck of medical image segmentation persists: MVSeg2023 provides only 105 annotated training volumes, an order of magnitude smaller than the datasets that power state-of-the-art segmentation frameworks such as nnU-Net in other anatomical domains~\cite{isensee2021nnunet}. Because 3D TEE annotation requires simultaneous expertise in cardiac anatomy and ultrasound interpretation, this scarcity is unlikely to be resolved through additional manual labelling alone.

A parallel line of work has explored synthetic data generation as a remedy for small medical imaging datasets. Generative Adversarial Networks (GANs) have been used to augment liver lesion classification~\cite{fridadar2018gan}, and surveys of GAN-based medical image synthesis document dozens of similar applications across radiology and pathology~\cite{kazeminia2020gans, yicomputingsurveys}. More recently, large-scale vision-language foundation models have been shown capable of synthesizing clinically coherent chest X-rays conditioned on radiology reports~\cite{bluethgen2025vision}, and a concurrent study demonstrated that a general-purpose commercial image generator (Nano Banana, the image-generation component of Google Gemini) can produce synthetic chest radiographs realistic enough to train a pneumonia classifier that generalizes to real-world X-ray datasets with an AUROC of up to $0.923$~\cite{peng2025pneumonia}. 

This study follows Peng et al.~\cite{peng2025pneumonia}, who showed that off-the-shelf, non-medical generative AI tools (requiring no fine-tuning or specialized medical training data) can produce synthetic images with measurable downstream classification utility. We ask whether this paradigm extends from 2D image \emph{classification} to 3D TEE \emph{segmentation}, a substantially harder problem for two reasons. First, segmentation requires a dense, voxel-accurate label for every training example, whereas classification requires a single scalar label, and general-purpose text-to-image generators such as ChatGPT, Gemini, and MAI-Image-2.6 cannot produce such masks. Second, 3D TEE is volumetric while consumer generative models output only 2D raster images, a dimensional mismatch that na\"ive augmentation cannot resolve.

We address both obstacles with a single procedure. Rather than treating synthetic images as independent training examples requiring annotation, we use each 2D image purely as a \emph{style} reference and apply Fourier Domain Adaptation (FDA)~\cite{yang2020fda} to transfer selected low-frequency amplitude components onto a real, already-annotated 3D TEE volume. Because FDA retains the real-image phase, the volume's spatial organization is expected to remain largely unchanged, so we reuse the original expert mask rather than claiming that every anatomical feature is exactly preserved. This turns the 2D-image-to-3D-mask annotation problem into a label-preserving appearance-transfer problem, yielding additional labelled 3D training volumes without new manual segmentation.

We construct three generator-guided FDA training sets, derived from ChatGPT (GPT-Image), Google Gemini (Nano Banana), and Microsoft MAI-Image-2.6, and compare them with a non-AI FDA control, a real-only baseline, and traditional augmentation, giving six experimental conditions. Our contributions are:

\begin{enumerate}

\item We evaluate general-purpose, non-medical generative-AI images as appearance-style donors for 3D TEE mitral valve segmentation.

\item We use slice-wise 2D FDA to bridge the dimensional mismatch between 2D generated images and 3D segmentation training: the synthetic image contributes low-frequency appearance, while the real-image phase and original expert mask are retained, avoiding voxel-level label generation.

\item We perform a controlled three-generator evaluation with a matched-update real-only baseline, a non-AI FDA style-donor control, a provenance-tracked FDA parameter sensitivity analysis, and case-level paired statistical testing across three runs.

\end{enumerate}

\section{Related Work}

\textit{Mitral Valve and Cardiac Structure Segmentation:}
Automated mitral valve segmentation from 3D TEE has progressed from atlas-based and active-contour methods to deep learning. Andreassen et al.\ used a deep convolutional network for mitral annulus segmentation in 3D TEE~\cite{andreassen2020annulus}. Carnahan et al.'s DeepMitral applied a 3D residual U-Net trained with the MONAI framework to segment the full leaflet apparatus directly from raw TEE volumes, achieving strong agreement with manual expert segmentations while reducing processing time from hours to seconds~\cite{carnahan2021deepmitral}. Subsequent work has introduced anatomical constraints into mitral annulus segmentation~\cite{anatomicalconstraints2024} and explored graph-neural-network representations of continuous leaflet motion across the cardiac cycle~\cite{mvgnn2022}. Chen et al.\ proposed an automatic 3D mitral valve leaflet segmentation pipeline validated against quantitative clinical measurements~\cite{chen2023mitralleaflet}. Across this literature, dataset size remains a recurring limitation, with reported training sets ranging from 44 to about 100 annotated volumes~\cite{carnahan2021deepmitral, regurgitantmv2024}---a scale far below that available for more common segmentation targets such as abdominal organs or brain tumors.

%\textit{General Medical Image Segmentation Architectures:} The 3D U-Net~\cite{cicek20163dunet}, extending the original 2D U-Net~\cite{ronneberger2015unet} to volumetric data with dense skip connections, remains a foundational architecture for medical segmentation. nnU-Net demonstrated that a largely automated, self-configuring adaptation of the U-Net family can match or exceed specialized architectures across 23 public biomedical segmentation benchmarks without manual tuning~\cite{isensee2021nnunet}, underscoring that architectural innovation alone has diminishing returns relative to appropriate data and preprocessing. 
%DeepMitral itself is built on MONAI, the PyTorch-based open-source medical imaging deep learning framework used throughout this study~\cite{cardoso2022monai}.

\textit{Synthetic Medical Image Generation and GAN-Based Augmentation:}
Frid-Adar et al.\ utilized GAN-generated synthetic CT liver lesions to improve CNN classification performance on a real dataset~\cite{fridadar2018gan}. Kazeminia et al.\ and Yi et al.\ subsequently surveyed the rapid expansion of GAN applications across medical image synthesis, segmentation, and cross-modality translation~\cite{kazeminia2020gans, yicomputingsurveys}. More recently, Bluethgen et al.\ demonstrated that a vision-language foundation model conditioned on radiology reports could synthesize chest radiographs with clinically coherent pathology~\cite{bluethgen2025vision}, illustrating the shift from adversarial architectures toward large pretrained generative foundation models for medical image synthesis.

\textit{General-purpose Generative AI for Medical Data Augmentation:}
Very recently, Peng et al.\ used Nano Banana (Google's general-purpose, non-medical-specialized image generation model) to synthesize 300 chest radiographs and fine-tuned a ResNet-50 classifier exclusively on this synthetic data, achieving an AUROC of $0.923$ on the real-world RSNA Pneumonia Detection dataset and $0.824$ on an independent Chest X-Ray dataset~\cite{peng2025pneumonia}. That study also compared against RoentGen-v2, a domain-specialized diffusion model for CXR generation~\cite{moroianu2025roentgen}, finding that the general-purpose Nano Banana model outperformed the CXR-specialized generator once post-processing (watermark and field-of-view cropping) was applied. 

%The present study is directly motivated by this result but targets a fundamentally different problem class: 3D volumetric segmentation rather than 2D binary classification, which as discussed in Section~I cannot use the same direct-generation strategy because segmentation requires dense spatial labels that a text-to-image model cannot supply.

\textit{Domain Adaptation and Fourier-Based Style Transfer:}
Unsupervised domain adaptation seeks to transfer a model trained on a source domain (e.g., synthetic images) to a target domain (e.g., real images) without target-domain labels, historically through adversarial feature alignment~\cite{hoffman2018cycada}.
Yang and Soatto proposed Fourier Domain Adaptation, showing that
swapping selected low-frequency amplitude components between source
and target images can alter appearance-related statistics while largely
retaining spatial organization~\cite{yang2020fda}. FDA and related
Fourier-based augmentation strategies have since been applied to
source-free medical image domain adaptation~\cite{yang2022fouriermining}.
%and general domain generalization~\cite{xu2021fourierframework}.
%We adapt this label-preserving property of FDA for a new purpose: rather than aligning a synthetic source domain to a real target domainfor inference, we use it to transfer appearance information from AI-generated donor images onto real, already-labelled training volumes.

%\textit{Self-Configuring and Robust Training Frameworks:}
%Beyond architecture and data, 

\section{Dataset and Baseline Model}

\textit{MVSeg2023 Dataset:}
We use the publicly released MVSeg2023 dataset~\cite{mvseg2023}, comprising 3D TEE volumes with voxel-level mitral valve leaflet annotations, partitioned into 105 training, 30 validation, and 40 test volumes. The dataset was originally released with separate anterior- and posterior-leaflet label classes. Following the pretrained DeepMitral model's convention, we merge these into a single binary leaflet class (background vs.\ leaflet) for all experiments, so that results are directly comparable to the publicly released pretrained checkpoint. Voxel spacing varies by volume (approximately $0.2$--$0.7$\,mm per axis), reflecting the heterogeneous acquisition protocols typical of clinical 3D TEE. The 40 held-out test volumes were used exclusively for evaluation and were not used to construct any FDA-adapted training volumes.

\textit{DeepMitral Model:}
We use the official DeepMitral implementation~\cite{carnahan2021deepmitral}, a 3D residual U-Net built on MONAI~\cite{cardoso2022monai} with five resolution levels (channel widths 16--32--64--128--256), stride-2 downsampling at each level, two residual units per level, batch normalization, and no dropout. 
Training uses random patch-based cropping (with a patch size of $96^3$ voxels, with foreground-biased sampling). 
%the NovoGrad optimizer with a learning rate of $1\times10^{-3}$, and a combined Dice--cross-entropy loss.

%%%

\section{Methodology}

\subsection{Overview}

\begin{figure*}[t]
\centering
\includegraphics[
    width=0.98\textwidth,
    keepaspectratio
]{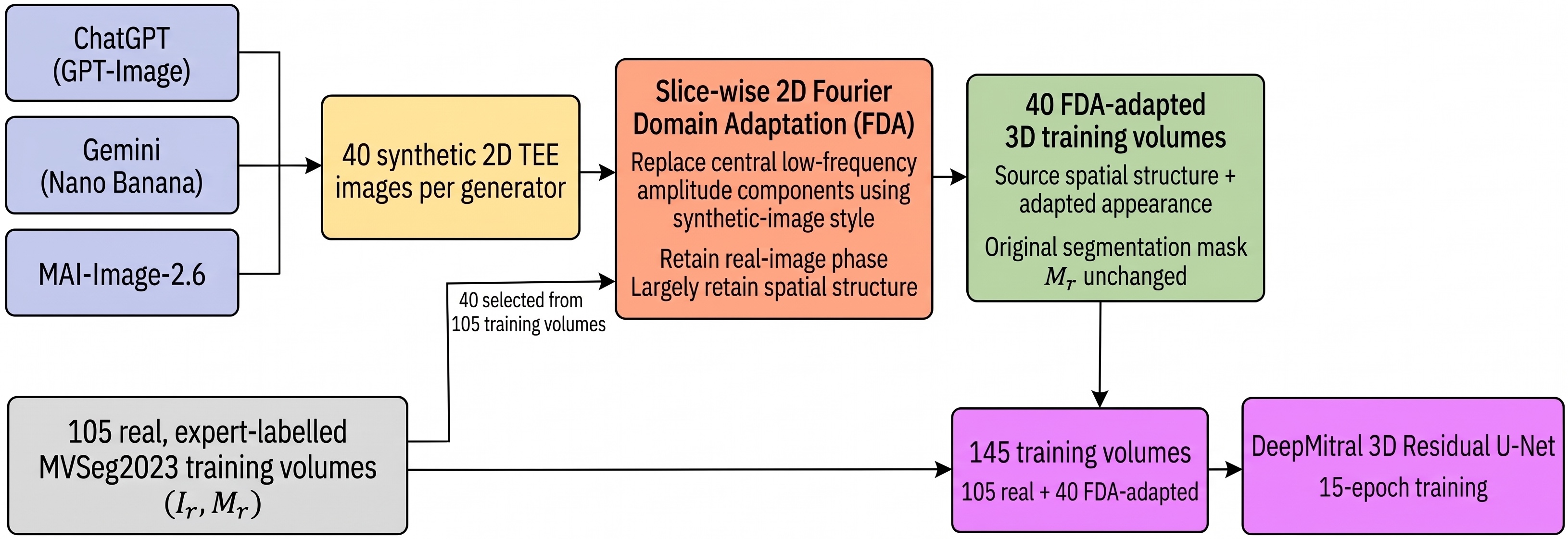}
\caption{
Overview of the generator-guided slice-wise 2D Fourier Domain Adaptation pipeline. For each generator, 40 synthetic 2D TEE images serve as appearance-style donors for 40 selected MVSeg2023 training volumes. FDA modifies selected low-frequency amplitude components while retaining the source-image phase. The original expert segmentation mask is reused unchanged. The 40 FDA-adapted volumes are combined with the 105 original real training volumes to form a 145-volume training set for DeepMitral.
}
\label{fig:pipeline}
\end{figure*}

Figure~\ref{fig:pipeline} summarizes the pipeline. For each of the three generative-AI tools, we (i) generate 40 synthetic 2D TEE images with structured prompts; (ii) apply slice-wise Fourier Domain Adaptation to transfer low-frequency appearance from each 2D donor onto a selected real training volume, retaining the real-image phase and reusing the original expert mask unchanged; (iii) combine the 40 adapted volumes with the 105 real training volumes to form a 145-volume training set; and (iv) train DeepMitral with an identical architecture and optimization protocol. The validation and test splits consist exclusively of real, expert-annotated volumes.

Six conditions are compared: a real-only baseline, traditional augmentation, and four FDA conditions (ChatGPT-, Gemini-, MAI-, and non-AI FDA). Because the FDA sets contain more volumes than the baseline and therefore receive more optimization updates, we additionally trained a 540-update real-only control matching their budget. A $\beta$-sensitivity analysis is also performed. 
%the non-AI FDA control and the $\beta$ sweep are described with their results in Sections~\ref{sec:nonai} and~\ref{sec:beta}.

\subsection{Synthetic Image Generation}

For each of three generators, ChatGPT (GPT-Image), Google Gemini (Nano Banana), and Microsoft MAI-Image-2.6, we generated 40 synthetic 2D images of the mid-esophageal TEE mitral valve view. Following the structured-prompt approach of Peng et al.~\cite{peng2025pneumonia}, an initial prompt requested a plain grayscale ultrasound sector image without on-image text, ECG traces, or angle indicators (which early generations contained), followed by class-specific prompts for normal morphology and for variants consistent with mitral regurgitation, stenosis, and prolapse, to promote structural diversity. All outputs were manually inspected, and images with encoding artifacts were discarded before augmentation.

\subsection{Slice-wise 2D Fourier Domain Adaptation}

Given a real 3D TEE volume $I_r$ with expert leaflet mask $M_r$ and a synthetic 2D image $I_s$, we applied conventional 2D FDA~\cite{yang2020fda} independently to each slice of $I_r$ (i.e., FDA was not implemented as a 3D Fourier transform). Each donor was paired with one real training volume and used for all of its slices. For each slice, the donor was resized to the slice dimensions, both images were normalized to $[0,1]$, and a 2D FFT was computed. The central low-frequency region of the slice's amplitude spectrum was replaced with the donor's corresponding components using $\beta=0.01$, while the slice's phase spectrum was retained. An inverse FFT reconstructed each adapted slice, and the slices were restacked in their original order to form the adapted volume $I_r^{\prime}$. Because the real phase is retained, spatial organization is expected to remain largely intact, so $M_r$ was reused unchanged: the donor contributes low-frequency appearance variation while supervision is inherited from the real volume. This lets generated 2D images add appearance diversity to 3D training without synthetic 3D masks or additional annotation.

For each generator, the 40 donors were paired with 40 distinct real training volumes, selected from the 105 with a fixed random seed of 42, which is independent of the training seeds (1, 2, and 3). The same selection and $\beta=0.01$ were used for all generators (i.e., $\beta$ was not tuned per generator).

\subsection{Training Configuration}

In order to realize a standardized, rigorously validated protocol, we held the network architecture, optimizer, patch size, and epoch-based training schedule fixed across all six experimental conditions, so that performance differences reflect the training data rather than the protocol. 
All experiments used the DeepMitral 3D residual U-Net with $96^3$-voxel patches, batch size 4, \texttt{drop\_last=True}, NovoGrad with learning rate $1\times10^{-3}$, the same Dice--cross-entropy loss, and a 15-epoch schedule, with three independent training runs per condition.
The primary real-only baseline (105 volumes) yields 26 mini-batches per epoch, i.e., 390 updates; each FDA condition (145 volumes) yields 36 mini-batches per epoch, i.e., 540 updates. The primary baseline used the 105 real training volumes without FDA or other synthetic augmentation, and its three runs were taken from archived original runs. The 540-update real-only control used the same 105 volumes, network, loss, optimizer, batch size, and three-run protocol. The validation and test volumes were never added to the training set.

The traditional-augmentation comparator used the 105 real volumes with online MONAI augmentations: \texttt{RandAffined} (probability $0.5$; rotations up to $\pm5^\circ$ per axis; scaling up to $\pm10\%$; bilinear interpolation for images, nearest-neighbor for masks; border padding), \texttt{Rand3DElasticd} (probability $0.2$; $\sigma\in[3,5]$; magnitude in $[1,2]$; same interpolation), and \texttt{RandShiftIntensityd} (image only; offset $0.10$; probability $0.5$). These parameters were fixed a priori and not tuned.

\subsection{Standardized Evaluation Metrics}

Segmentation quality is assessed with the Generalized Dice Score (GDS) for overlap, the 95th-percentile Hausdorff Distance (HD95) for boundary discrepancy, and Surface Distance for mean surface error. Since the anterior and posterior leaflets are merged into one foreground class, GDS reduces to the standard Dice coefficient; we retain the term GDS to reflect the MONAI metric used. All test metrics were computed with an identical protocol: predictions were defined on the DeepMitral 0.3-mm isotropic evaluation grid, the ground truth was resampled to the prediction shape and affine with nearest-neighbor interpolation, and all three metrics were computed on the binary foreground with background excluded. Distances were converted from voxels to mm using the isotropic spacing.

%%%

\section{Results}

\subsection{Statistical Analysis Methods}
\label{sec:stats}

This subsection describes how the reported numbers and $p$-values were obtained, for the ease of the reader.

\textit{Case-level aggregation.} Model performance is reported as mean $\pm$ sample standard deviation across three independent training runs. The public MVSeg2023 release provides anonymized, single-frame end-diastolic volumes whose metadata are stripped except for image size and spacing; the dataset description states that each case represents one patient, but no patient identifiers or clinical attributes are released. The unit of analysis is therefore the test case (volume), and independence across cases is assumed from the dataset description rather than verified. Because all runs are evaluated on the same 40 held-out cases, conditions are compared at the case level: each metric was first averaged across the three runs for every case and condition, giving one value per case and condition ($n=40$) that uses all runs rather than a single seed.

\textit{Paired comparisons.} For two conditions, the paired difference is computed for each case and defined so that positive values indicate improvement (GDS increase; HD95 and Surface Distance reduction). Significance is assessed with a two-sided Wilcoxon signed-rank test, a non-parametric test of whether the paired differences are centered on zero that does not assume normally distributed differences ($W$ denotes its test statistic). Uncertainty in the mean paired difference is given by a paired bootstrap 95\% confidence interval (CI), obtained by resampling cases with replacement. A CI that contains zero means the mean difference is not clearly distinguishable from zero; this can occur alongside a significant Wilcoxon result, because the rank-based test and the mean-based interval summarize the difference differently.

\textit{Multiple comparisons.} Because three metrics are tested per comparison, Holm correction was applied across the three metrics within each pre-specified comparison, and all $p$-values reported below are Holm-adjusted ($p_{\mathrm{Holm}}$) unless stated otherwise. As a conservative robustness check, Holm correction was also applied jointly across all nine generator-by-metric comparisons against the 390-update baseline ($p_{\mathrm{Holm,9}}$). The comparisons performed were: ChatGPT-, Gemini-, and MAI-FDA against the 390-update real-only baseline; ChatGPT-FDA and the non-AI FDA control against the 540-update real-only control; and the non-AI FDA control against ChatGPT-FDA.

\textit{FDA Parameter sensitivity.} For the FDA parameter $\beta$ sensitivity experiment, case-level metrics were averaged over runs for each $\beta$ and compared with an overall Friedman test, a non-parametric test for differences across three or more related conditions, with Holm-corrected pairwise Wilcoxon tests and paired bootstrap intervals for follow-up. The Friedman analysis was repeated after excluding Seed~1 (averaging Seeds~2 and~3) to assess sensitivity to between-run variability. Friedman $p$-values are reported as obtained.

%\textbf{Reporting.} Mean differences are computed from unrounded case-level values and may differ in the last digit from differences of the rounded means in Table~\ref{tab:results}.

\subsection{Quantitative Segmentation Performance}
\label{sec:quant}

\begin{table*}[t]
\centering
\caption{
Standardized three-run segmentation performance on the held-out MVSeg2023 test set. Results are mean $\pm$ sample standard deviation across three independent runs. GDS was computed on the foreground leaflet class with background excluded. Higher GDS is better; lower HD95 and Surface Distance are better. Bold indicates the numerically best mean among the six conditions and does not imply statistical superiority.
}
\label{tab:results}

\resizebox{0.88\textwidth}{!}{%
\begin{tabular}{lccc}
\toprule
\textbf{Condition}
& \textbf{GDS $\uparrow$}
& \textbf{HD95 (mm) $\downarrow$}
& \textbf{Surface Distance (mm) $\downarrow$} \\
\midrule

Real-only baseline
& 0.8198 $\pm$ 0.0117
& 6.0005 $\pm$ 1.6528
& 1.2911 $\pm$ 0.2886 \\

Traditional augmentation
& 0.8077 $\pm$ 0.0175
& 6.2557 $\pm$ 2.4475
& 1.3235 $\pm$ 0.6363 \\

Real + ChatGPT-FDA
& 0.8268 $\pm$ 0.0307
& 3.0569 $\pm$ 0.1696
& 0.5866 $\pm$ 0.1098 \\

Real + Gemini-FDA
& 0.8192 $\pm$ 0.0291
& 3.0816 $\pm$ 0.5196
& \textbf{0.5712 $\pm$ 0.1460} \\

Real + MAI-FDA
& \textbf{0.8309 $\pm$ 0.0129}
& 3.6890 $\pm$ 1.0765
& 0.7028 $\pm$ 0.2258 \\

Real + Non-AI FDA
& 0.8259 $\pm$ 0.0268
& \textbf{3.0208 $\pm$ 0.3519}
& 0.5824 $\pm$ 0.0765 \\

\bottomrule
\end{tabular}%
}
\end{table*}

Table~\ref{tab:results} reports the standardized test performance of all six conditions. The four FDA-based conditions (ChatGPT-, Gemini-, MAI-, and Non-AI FDA) substantially reduced mean HD95 and Surface Distance relative to the real-only baseline, while their overlap results differed. {MAI-FDA had the numerically highest mean GDS (0.8309 $\pm$ 0.0129), the non-AI FDA control the numerically lowest mean HD95 (3.0208 $\pm$ 0.3519~mm, narrowly below ChatGPT-FDA at 3.0569 $\pm$ 0.1696~mm), and Gemini-FDA the numerically lowest mean Surface Distance (0.5712 $\pm$ 0.1460~mm).} Traditional augmentation yielded lower mean GDS and higher mean boundary errors than the baseline. These orderings are descriptive; the corresponding paired tests are reported in Sections~\ref{sec:paired} and~\ref{sec:nonai}.

\subsection{Paired Case-Level Analysis}
\label{sec:paired}

Paired results are summarized in Table~\ref{tab:paired}, using the conventions of Section~\ref{sec:stats}.

\begin{table*}[t]
\centering
\caption{
Case-level paired comparisons ($n=40$; metrics averaged over three runs). Entries are mean paired improvements (positive = better) with Holm-adjusted two-sided Wilcoxon $p$-values in parentheses. Holm correction was applied across the three metrics within each comparison. A dash indicates that the mean difference is not reported; the corresponding $p$-value is given.
}
\label{tab:paired}
\resizebox{0.95\textwidth}{!}{%
\begin{tabular}{llccc}
\toprule
\textbf{Reference} & \textbf{Condition}
& \textbf{$\Delta$GDS ($p_{\mathrm{Holm}}$)}
& \textbf{$\Delta$HD95, mm ($p_{\mathrm{Holm}}$)}
& \textbf{$\Delta$Surface Dist., mm ($p_{\mathrm{Holm}}$)} \\
\midrule
390-update real-only & ChatGPT-FDA & $+0.0071$ ($0.166$) & $2.9435$ ($3.41\times10^{-4}$) & $0.7046$ ($5.46\times10^{-12}$) \\
390-update real-only & Gemini-FDA  & $-0.0005$ ($0.837$) & $2.9189$ ($0.0117$) & $0.7199$ ($1.36\times10^{-10}$) \\
390-update real-only & MAI-FDA     & $+0.0111$ ($0.00123$) & $2.3115$ ($5.69\times10^{-5}$) & $0.5884$ ($9.49\times10^{-9}$) \\
\midrule
540-update real-only & ChatGPT-FDA & $+0.0347$ ($5.46\times10^{-12}$) & $0.7020$ ($8.04\times10^{-4}$) & $0.0663$ ($0.0321$) \\
540-update real-only & Non-AI FDA  & $+0.0337$ ($5.46\times10^{-12}$) & $0.7382$ ($7.58\times10^{-4}$) & --- ($0.0603$) \\
\bottomrule
\end{tabular}%
}
\end{table*}

\textbf{Comparison with the 390-update baseline.} ChatGPT-FDA increased mean GDS from 0.8198 to 0.8268, a non-significant paired improvement of 0.0071 (bootstrap 95\% CI: $[-0.0011,\,0.0168]$; $W=306$), with higher GDS in 23 of 40 cases and lower in 17. Gemini-FDA likewise showed no GDS difference ($-0.0005$; 95\% CI: $[-0.0106,\,0.0102]$; $W=394$; higher in 20 cases, lower in 20). In contrast, MAI-FDA significantly improved GDS by 0.0111 (95\% CI: $[0.0034,\,0.0209]$; $W=176$), with improvement in 30 of 40 cases. HD95 and Surface Distance were significantly reduced for all three generator-guided conditions; for ChatGPT-FDA, HD95 decreased from 6.0005 to 3.0569~mm (95\% CI: $[1.2460,\,5.1259]$~mm; improved in 28 of 40 cases) and Surface Distance from 1.2911 to 0.5866~mm (95\% CI: $[0.3337,\,1.2319]$~mm; improved in all 40). The boundary improvements were therefore consistent across generators, whereas the GDS effect was generator-dependent. These conclusions were unchanged under a more conservative Holm correction applied jointly across all nine generator-by-metric tests: the MAI-FDA GDS improvement remained significant ($p_{\mathrm{Holm,9}}=0.00492$), as did all HD95 and Surface Distance reductions, and the ChatGPT-FDA and Gemini-FDA GDS differences remained non-significant.

\textbf{Comparison with the 540-update control.} The optimization-matched real-only control achieved GDS of $0.7921\pm0.0441$, HD95 of $3.7590\pm1.1742$~mm, and Surface Distance of $0.6529\pm0.2691$~mm. Relative to it, ChatGPT-FDA increased GDS by 0.0347 (95\% CI: $[0.0266,\,0.0439]$; $W=0$), with higher GDS in all 40 cases. HD95 and Surface Distance were also lower (Table~\ref{tab:paired}), but the bootstrap confidence intervals for these two mean reductions crossed zero. The two baseline comparisons answer complementary questions: the 390-update comparison tests ChatGPT-FDA against the primary real-only condition in Table~\ref{tab:results}, whereas the 540-update comparison controls for the larger optimization budget of the FDA conditions.

\subsection{Non-AI FDA Style-Donor Control}
\label{sec:nonai}

To test whether the effect required generative-AI donor content, we constructed a non-AI FDA control in which the style donors were real TEE images drawn exclusively from the MVSeg2023 training partition (no validation or test image was used as a donor). The same 40 real source volumes and $\beta=0.01$ were used; the 40 adapted volumes were combined with the original 105 training volumes, giving the same 145-volume, 540-update budget as the generator-guided conditions, and training was repeated over three independent runs.

{Its three-run mean performance is reported in Table~\ref{tab:results}.} Relative to the 540-update real-only baseline, GDS increased by 0.0337 (95\% CI: $[0.0262,\,0.0423]$; $p_{\mathrm{Holm}}=5.46\times10^{-12}$), and HD95 was lower by 0.7382~mm ($p_{\mathrm{Holm}}=7.58\times10^{-4}$), although the bootstrap confidence interval for the mean HD95 reduction crossed zero. The Surface Distance difference was not significant ($p_{\mathrm{Holm}}=0.0603$). Compared directly with ChatGPT-FDA, no significant difference was detected for GDS ($p_{\mathrm{Holm}}=0.628$) or HD95 ($p_{\mathrm{Holm}}=0.628$); Surface Distance showed a paired-rank difference ($p_{\mathrm{Holm}}=0.0293$), but its bootstrap confidence interval crossed zero. {Thus, although the non-AI control attains the numerically lowest mean HD95 in Table~\ref{tab:results}, this is not a statistically significant advantage over ChatGPT-FDA.} These experiments do not establish that generative-AI donor content is superior to real-image donor content; they support FDA itself as an important component of the observed augmentation effect.

\subsection{FDA Parameter Sensitivity Analysis}
\label{sec:beta}

We performed a provenance-tracked $\beta$-sensitivity experiment for ChatGPT-FDA with $\beta\in\{0.005,0.010,0.020\}$. The same 40 source cases, 40 ChatGPT donor images, donor assignments, and pairing order were held fixed; only the Fourier low-frequency replacement parameter changed. Three deterministic 540-update runs were performed per setting. Three-run means are given in Table~\ref{tab:beta}.

\begin{table}[t]
\centering
\caption{$\beta$-sensitivity results for ChatGPT-FDA (mean $\pm$ sample standard deviation over three runs).}
\label{tab:beta}
\resizebox{\columnwidth}{!}{%
\begin{tabular}{lccc}
\toprule
$\beta$ & GDS $\uparrow$ & HD95 (mm) $\downarrow$ & Surface Dist. (mm) $\downarrow$ \\
\midrule
0.005 & $0.8133\pm0.0523$ & $3.5530\pm1.2649$ & $0.6131\pm0.1972$ \\
0.010 & $0.8082\pm0.0611$ & $3.4409\pm1.1327$ & $0.5688\pm0.0686$ \\
0.020 & $0.8295\pm0.0248$ & $3.1804\pm0.3042$ & $0.6177\pm0.0540$ \\
\bottomrule
\end{tabular}%
}
\end{table}

Using case-level metrics averaged over all three runs, the Friedman test indicated differences across $\beta$ for GDS ($\chi^2(2)=52.55$, $p=3.88\times10^{-12}$), HD95 ($\chi^2(2)=29.85$, $p=3.30\times10^{-7}$), and Surface Distance ($\chi^2(2)=18.60$, $p=9.14\times10^{-5}$). This sensitivity was not robust to exclusion of Seed~1: using Seeds~2 and~3 only, the Friedman test was non-significant for GDS ($p=0.741$), HD95 ($p=0.332$), and Surface Distance ($p=0.054$). The full-run differences are therefore attributed largely to between-run variability rather than to a consistently optimal $\beta$.

%\subsection{Reproducibility and Provenance Audit}

%The regenerated $\beta=0.010$ condition did not reproduce the absolute performance of the initial ChatGPT-FDA experiment (Table~\ref{tab:results}) despite deterministic reproducibility within the regenerated pipeline. Because the intermediate FDA volumes of the initial experiment were not archived, the source of this difference could not be resolved retrospectively; the $\beta$ sweep is therefore treated as an internally controlled analysis, and comparisons are made only among the three regenerated $\beta$ conditions.

%As a separate audit, we reran inference on the 40-case test set using the preserved original ChatGPT-FDA Seed~3 540-update checkpoint and the archived DeepMitral implementation, with the same binary label normalization ($2\rightarrow1$) as the evaluation pipeline. All 40 reproduced predictions were identical to the recovered Seed~3 predictions at the compressed-file, decompressed NIfTI, voxel-array, affine, and image-shape levels, with zero differing voxels. This establishes checkpoint-level reproducibility of the recovered predictions, not identity with a separate archive of the originally written prediction volumes (none was retained).

%%%

\section{Discussion}

The standardized three-run analysis shows that FDA-based augmentation does not produce a uniform improvement across every segmentation metric. Among the six experimental conditions reported in Table~\ref{tab:results}, MAI-FDA had the numerically highest mean GDS, the non-AI FDA control had the numerically lowest mean HD95, and Gemini-FDA had the numerically lowest mean Surface Distance. All four FDA-based conditions substantially reduced mean boundary-error metrics relative to the primary real-only baseline, whereas traditional augmentation did not improve the primary baseline. However, since the parameters of traditional augmentation were fixed a priori and not tuned, one should not conclude that the conventional augmentation was ineffective.

The control experiments further refine the interpretation of these results. 
Relative to the 390-update baseline, the GDS effect was donor-dependent: only MAI-FDA gave a significant paired improvement ($\Delta=0.0111$, $p_{\mathrm{Holm}}=0.00123$), whereas HD95 and Surface Distance were significantly reduced for all three generator-guided conditions. The most reproducible effect is thus reduced boundary and surface error, while improved overlap is not shared across generators.
%The present experiments establish this statistical heterogeneity but do not identify which specific donor-image characteristics account for the different GDS responses.

The 540-update real-only control addresses optimization budget. ChatGPT-FDA showed a significant GDS advantage over it, so its result is not attributable solely to extra updates. Because this control had lower GDS but better boundary metrics than the 390-update baseline, longer training appears to shift the overlap--boundary trade-off. The two controls are therefore treated as complementary. 

Real training-image style donors yielded GDS and HD95 similar to ChatGPT-FDA (no significant difference), so FDA is an important mechanism but generative-AI donor content is not shown to be uniquely beneficial. Its practical value is as an annotation-free source of 2D appearance variation transferable to labelled real volumes, rather than a generator of valid 3D volumes or labels. 
In other words, generative-AI donor may be particularly useful when additional real appearance donors are unavailable or costly to collect.

Finally, The FDA parameter $\beta$ sensitivity analysis therefore points to substantial between-run variability and does not support a universal claim that one tested $\beta$ value is optimal.

%%%

\textbf{Limitations.} First, MVSeg2023 is anonymized, with no patient identifiers or diagnostic metadata, so independence across cases is assumed and performance could not be stratified by valve pathology. Also, the test cohort has only 40 real MVSeg2023 volumes. Thus, external validation on an independent 3D TEE dataset is needed. Second, FDA alters appearance statistics but creates no new valve anatomy or pathology. Third, only three runs per condition limit inference on between-run variability. Fourth, clinical morphometrics (e.g., mitral annular area, inter-commissural distance, tenting area) could not be derived because the binary masks lack the required landmarks and reference planes. Fifth, the six-condition comparison reported in Table~\ref{tab:results} used an epoch-matched design (390 updates for the primary real-only baseline, 540 for FDA conditions). 
We therefore report both a direct paired comparison with the original 390-update Table~1 baseline and a separate 540-update real-only control that matches the FDA optimization budget. The GDS conclusions differ between these comparisons: the ChatGPT-FDA GDS increase was not significant relative to the original 390-update baseline but was significant relative to the 540-update matched control. Accordingly, overlap and boundary metrics are interpreted separately rather than treating either baseline comparison as universally definitive.
%We therefore report both baseline comparisons, whose GDS conclusions differ for ChatGPT-FDA, and interpret overlap and boundary metrics separately. 
%Sixth, the intermediate FDA volumes of the initial ChatGPT-FDA experiment were not archived, so the regenerated $\beta=0.010$ condition does not reproduce the historical construction and the $\beta$ sweep is an internally controlled analysis. However, the preserved Seed~3 checkpoint and DeepMitral code reproduced all 40 recovered test predictions exactly, verifying prediction-level reproducibility but not the provenance of the unarchived volumes.

\section{Conclusion}

We evaluated general-purpose generative-AI images as 2D appearance donors for label-preserving, slice-wise Fourier Domain Adaptation in 3D mitral-valve segmentation. In the standardized six-condition three-run comparison, MAI-FDA achieved the numerically highest mean GDS (0.8309 $\pm$ 0.0129), the non-AI FDA control the lowest mean HD95 (3.0208 $\pm$ 0.3519 mm), and Gemini-FDA the lowest mean Surface Distance (0.5712 $\pm$ 0.1460 mm). Relative to the 390-update real-only baseline, HD95 and Surface Distance were significantly reduced for all three generative-AI donors, making improved boundary accuracy the most consistent finding, whereas the GDS effect was generator-dependent (significant only for MAI-FDA: $\Delta=0.0111$, $p_{\mathrm{Holm}}=0.00123$), although ChatGPT-FDA also showed a significant GDS advantage over the 540-update matched control. Since a non-AI FDA control using real training images as style donors performed comparably on GDS and HD95, this supports FDA as an important augmentation mechanism without establishing that generative-AI donors are superior to real-image donors. Generative models nonetheless offer a practical, annotation-free source of 2D appearance diversity that can be transferred to already-labelled real 3D volumes without requiring synthetic voxel-level masks.

\section*{Appendix}

A common structured prompt template was used to generate synthetic 2D TEE images with ChatGPT (GPT-Image), Google Gemini (Nano Banana), and Microsoft MAI-Image-2.6:

{\small
\noindent
\textit{
Generate 10 realistic synthetic transesophageal echocardiography (TEE)
grayscale B-mode ultrasound images of the mitral valve, representing
10 different cases, generated individually as separate images.
Use a true mid-esophageal TEE mitral-valve view rather than a
transthoracic apical four-chamber view. The left atrium should appear
in the near field, the mitral valve should be prominently centered,
and the left ventricle should appear deeper in the ultrasound sector.
Focus mainly on the left atrium, mitral valve, and proximal left
ventricle. Clearly depict two separate mitral leaflets: the anterior mitral
leaflet (AML) and posterior mitral leaflet (PML), with the AML generally
longer than the PML. Both leaflets should arise from separate annular
attachment points, with visible leaflet bodies, free edges, and tips. Use a realistic fan-shaped grayscale B-mode ultrasound sector on a
black background with natural speckle noise and realistic echogenicity.
Introduce variation among cases in age, sex, cardiac phase, probe
angle, sector depth, image gain, speckle pattern, mild acoustic
shadowing, and heart size. Do not include any text, anatomical labels, letters, numbers, ECG
traces, color Doppler, depth markers, angle indicators, machine
interface elements, logos, or watermarks. Do not depict the mitral
valve as a single continuous bright horizontal line.
}
}

For normal mitral-valve images, the prompt additionally specified thin,
pliable, normally mobile leaflets with complete coaptation and no
thickening, calcification, stenosis, or prolapse. For mitral
regurgitation, the prompt requested incomplete leaflet coaptation with
a visible gap, redundant or thickened leaflet tissue, and/or a
prolapsing leaflet segment without color Doppler. For mitral stenosis,
the prompt requested thickened and/or calcified leaflets, restricted
diastolic opening, reduced leaflet separation, leaflet doming, and a
narrowed mitral orifice. For mitral valve prolapse, the prompt requested
one or both leaflets billowing or prolapsing toward the left atrium
during systole. The same core anatomical and imaging constraints were used with
ChatGPT, Gemini, and MAI-Image-2.6. Minor wording adjustments were made
during iterative generation when necessary to remove unwanted text,
interface elements, incorrect viewing geometry, or other non-diagnostic
artifacts.

%\section*{Acknowledgment}
%The authors thank the developers of the DeepMitral architecture and the MVSeg2023 dataset for making their code and data publicly available, and the MONAI Consortium for the open-source deep learning framework used throughout this work.

\end{document}